\documentclass{article}

\usepackage{amsmath,amsfonts,latexsym,amssymb,amsthm,graphicx,natbib}

\newcommand{\xbold}{ {\mathbf x}}
\newcommand{\bbold}{ {\mathbf b}}
\newcommand{\sbold}{ {\mathbf s}}
\newcommand{\tbold}{ {\mathbf t}}
\newcommand{\phibold}{\mbox{\boldmath${\phi}$}}
\newcommand{\psibold}{\mbox{\boldmath${\psi}$}}

\def\argmin{\mathop{\rm{argmin}}}

\title{Maximal Autocorrelation Functions in Functional Data Analysis}
\author{Giles Hooker\footnote{Department of Biological Statistics and Computational Biology, Cornell University, gjh27@cornell.edu} \  and Steven Roberts\footnote{Research School of Finance, Actuarial Studies and Applied Statistics, Australian National University, steven.roberts@anu.edu.au}}

\date{}

\begin{document}

\maketitle

\begin{abstract}
This paper proposes a new factor rotation for the context of functional principal components analysis.
This rotation seeks to re-represent a functional subspace in terms of directions of decreasing smoothness as represented by a generalized smoothing metric. The rotation can be implemented simply and we show on two examples that this rotation can improve the interpretability of the leading components.
\end{abstract}


\section{Introduction}

This paper proposes a new factor rotation for functional principal components analysis (fPCA). In functional data analysis, the use of principal components has received considerable attention; means of defining principal components were studied in \citet{RiceSilverman91} and \citet{Silverman96} and for sparsely observed curves in \citet{YMW05a,PengPaul09} and \citet{PengPaul11}.  Following carrying out fPCA, one of the most common means of dealing with functional covariates has been by employing principal components scores within multivariate methods. Particular examples include linear models \citep{YMW05b,HMW06,GBCCR11}, as responses \citep{CMW04,SenturkMuller06} and additive models \citep{MullerYao08}. In some of these cases, interpretation is gained by combining coefficients from the multivariate model with the principal component functions to create a functional parameter -- see the functional linear models in \citet{YMW05b} -- but this is not always possible, as in the additive models in \citet{MullerYao08}, in which case the model must be interpreted by treating the principal component directions as having particular meanings.

Despite this interest in fPCA, little has been proposed by the way of factor rotations that might make principal components directions more interpretable. \citet{RamsaySilverman05} examine an extension of the VARIMAX rotation \citep{Kaiser58} from multivariate factor analysis which has the tendency to produce components that focus on particular ranges of the domain of the functions. \citet{LRHF12} propose a rotation towards periodic components in a remote sensing example with functions that cover multiple years with a distinct annual signal.  Other methods from the multivariate factor rotation literature could be considered, but we have found no other suggested factor rotations that make use of the structure of functional data. In this paper, we propose a rotation towards maximally smooth principal components. These are the directions in which there is greatest predictability over time and which are also more interpretable.


The factor rotation that we propose is derived from the definition of Min/Max Autocorrelation factors (MAF) introduced by
\citet{switzer1989min} and \citet{switzer1984min} for the analysis of gridded multivariate data and parallel time series data, respectively. The principal underlying MAF is to find linear combinations of the original data that have maximum autocorrelation. This property of MAF is in contrast to PCA which finds linear combinations that have maximum variance. Of particular importance for our setting is the fact that when applied to parallel time series a MAF analysis finds linear combinations of the data that are decreasingly smooth functions of time or, in other words, that are decreasingly predictable functions of time.  In this vein, we show that a functional analogue of MAF can be obtained that searches for the rotated components that have smallest integrated first derivative. We then demonstrate how this can be extended to any notion of smoothness as given by a linear differential operator defined in \citet{RamsaySilverman05}. In our examples, we have developed our methods based on the numerical machinery in the {\tt fda} package in {\tt R} \citep[see][]{RamsayHookerGraves} but they can be readily employed with alternative functional data representations.

In recent literature, interest in the application and theoretical properties of MAF has been increasing \citep{cunningham2014unifying,CEM:CEM2634}. Of particular relevance to the current work is the paper of \citet{Henderson08} that compares PCA and MAF in the context of ToF-SIMS image interpretation. The authors conclude that MAF is more effective than PCA for the analysis of high signal intensity data. The importance of MAF for forecasting has been investigated by \citet{Woillez} in the context of fish stocks. The utility of MAF for forecasting correlated functional data is ongoing work.

The remainder of the paper is structured as follows: we derive the notion of maximally smooth rotation as an analogue of maximally autocorrelated time series in Section \ref{sec:MAF}. The extension of these to any linear differential operator and its implementation in basis expansion methods is given in Section \ref{sec:Rotation} and we demonstrate the effect of these methods in Section \ref{sec:data}. We finish with some concluding remarks and further directions.

\section{Maximal Autocorrelation Factor Rotations (MAFR)} \label{sec:MAF}

Our methods are developed on top of the Maximal Autcorrelation Functions proposed in \citet{switzer1984min} for multivariate time series. Suppose that we have a multivariate time series $\xbold_1,\ldots,\xbold_T$ in which at each time point the observed $\xbold_t$ is a vector. The maximally autocorrelated time series is the linear transformation $\bbold$ such that $\mbox{cor}(\xbold_t \bbold, \xbold_{t+1} \bbold)$ is maximized. In order to apply this to functional data analysis, we re-interpret the criterion as
\[
\hat{\bbold} = \argmin_{\bbold} \frac{ \sum_{t=1}^{T-1} \left( \bbold^T \xbold_{t+1} - \bbold \xbold_t \right)^2}{ \bbold^T \sum_{t=1}^T \xbold_t\xbold_t^T \bbold }.
\]
In a functional data analysis context we consider $\xbold_t$ to derive from the evaluation of a vector of functions $\xbold(t)$ at times $t = i(\Delta t)$ for $i = 1,\ldots,T$. By dividing by $\Delta t$ we can re-represent this criterion as
\[
\hat{\bbold} = \argmin_{\bbold} \frac{  \int \bbold^T \dot{\xbold}(t) \dot{\xbold}(t)^T  \bbold dt }{ \bbold^T \int \xbold(t) \xbold(t)^T dt \bbold }.
\]
where $\dot{\xbold}(t)$ is the vector of time-derivatives of $\xbold(t)$.

Here, we recognize the numerator as having the form of a classical first-derivative smoothing penalty on the univariate function $z(t) = \bbold^T \xbold(t)$. In this spirit, we can more generally define a criterion by any linear differential smoothing operator $L$ as in \citet{RamsaySilverman05}. This allows us to define the MAFR criterion as
\[
\mbox{MAFR}_L(\bbold) =  \frac{  \int \bbold^T L \xbold(t) L\xbold(t)^T  \bbold dt }{ \bbold^T \int \xbold(t) \xbold(t)^T dt \bbold }
\]
where the operator $L$ is a linear combination of derivatives:
\[
L \xbold(t) = \frac{d^k }{dt^k} \xbold_t + \sum_{j=0}^{k=1} a_j(t) \frac{d^j}{dt^j} \xbold(t).
\]
The most common choices for $L$ correspond to the first and second derivatives: $L \xbold(t) = \dot{\xbold}(t)$ or $L \xbold(t) = \ddot{\xbold}(t)$ but more complex penalties can also be useful and we examine the {\em harmonic acceleration} penalty
\[
L \xbold(t) = \frac{d^3}{dt^3} \xbold(t) - \frac{\omega}{2 \pi} \frac{d}{dt} \xbold(t)
\]
which defines smoothness in terms of sin and cosin functions with period $\omega$ as well as constant shifts \citep[see][]{RamsaySilverman05}.

We have written our criterion in terms of a collection of functions $\xbold(t)$ above, but this method is treated as a factor rotation to be applied following fPCA with a fixed number of components selected. Thus, below we will replace $\xbold$ with $\phibold(t) = (\phi_1(t),\ldots,\phi_K(t))$ to conform to more common notation.  If the dimension of $\xbold(t)$ is allowed to grow, we will always be able to reduce MAFR by adding further components, yielding rotations in which $L \bbold \xbold \rightarrow 0$. This same phenomenon occurs for classical factor rotations in multivariate analysis when the number of variables increases or in MAFs with an increasing number of time series. \citet{LRHF12} found that applying VARIMAX rotations to a large number of principal components resulted in essentially uninterpretable results. Similar comments may be made about the maximal autocorrelation functions in \citet{switzer1984min}.  Along similar lines, we expect that the trailing components after rotation will be the least interpretable and there is a trade-off between increasing the smoothness of the leading components and allowing some variation to be absorbed into the remaining, less-interpretable versions. These comments also apply to other factor rotations, although in the examples below we find that the leading components are smoothed while the remaining ones are relatively unaffected.


\section{Numerical Implementation} \label{sec:Rotation}

In this section, we describe the numerical implementation of the factor rotation. This can be accomplished easily using the basis expansion methods in the {\tt fda} package in {\tt R} \citep{RWGH13,RamsayHookerGraves}, but it relies only on our ability to obtain inner products of the derivatives of principal component functions.

We assume that a set of principal components $\phibold(t)$ have been obtained from data. Since these are orthonormal by definition, we have $\int \phibold(t)^T \phibold(t) dt$ is given by the identiy, and thus the MAFR rotation corresponds to
\[
\hat{\bbold} = \argmin_{\bbold} \bbold^T \left[ \int L\phibold(t)^T L\phibold(t) dt \right] \bbold, \mbox{ subject to } \bbold^T \bbold = 1
\]
By standard arguments, the solution to this problem is the smallest eigenvector of the matrix
\[
P = \left[ \int L\phibold(t)^T L\phibold(t) dt \right].
\]
We may define successive rotations $\bbold_2,\ldots,\bbold_k$ by minimizing MAFR($\bbold$) subject to $\bbold_i^T \bbold_j = I_{i=j}$. These are given by the succeeding columns of $U$ in the Eigen-decomposition
\[
P = UDU^T.
\]
We can thus define a rotation to new components
\[
\psibold(t) = U^T \phibold(t)
\]
If, as is standard,  the diagonal matrix $D$ is ordered from largest to smallest eigenvalues, the final components of $\psibold$ should be the smoothest. We observe that since both $U$ and the $\phibold$ are orthonormal, so are the $\psibold$.

If we have an original set of curves represented in terms of principal component scores
\[
x_i(t) = \sum_{k=1}^K s_{ij} \phi_j(t) = \sbold_i^T \phibold(t)
\]
the score vector $\sbold_i$ can be re-represented in the basis defined by $\psibold(t)$ in terms of $\tbold_i = U^T \sbold_i$. If the variances of the original retained principal components are given in the diagonal matrix $\Sigma$, the MAFR components have scores with associated covariance $U^T \Sigma U$.


\section{Examples} \label{sec:data}

\subsection{A Simulated Experiment} \label{sec:simulated}
We begin by experimenting with the effect of this rotation on simulated data in which rotation should help to capture a ``true'' set of leading principal component directions. For this simulation we represented 100 curves via a Fourier basis expansion with bases
\[
    f_0(t) = 1, \ f_{2i}(t) = \sin(2 \pi i t), \ f_{2i+1}(t) = \cos(2 \pi i t)
\]
and simulated the coefficients of the first 25 such basis functions as independent normals with exponentially decreasing variance:
\[
x_i(t) = \sum_{j = 0}^{24} c_j f_j(t), \ c_j \sim N(0, \exp(-j/4)).
\]
Under this framework, 10 principal components are required to capture 99\% of the variation in these data.

\begin{figure}
\begin{center}
\begin{tabular}{cc}
\includegraphics[height=4cm]{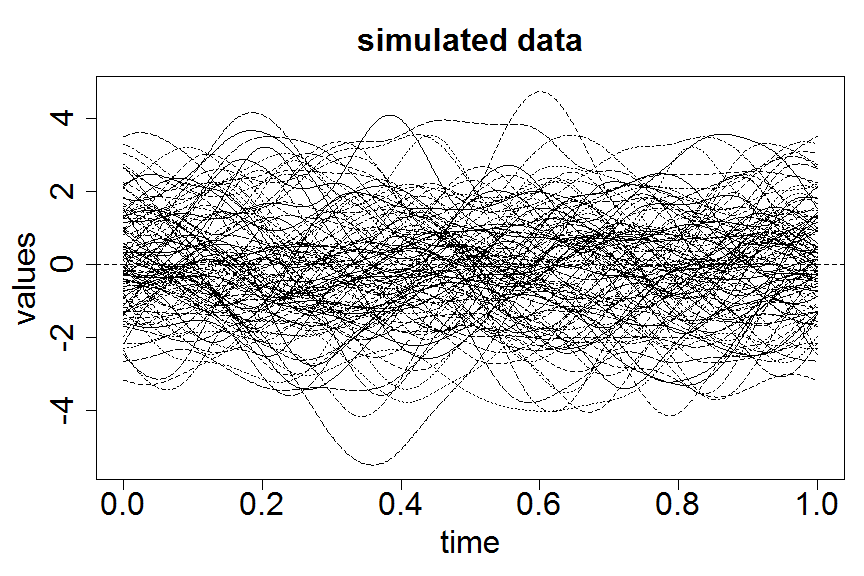} &
\includegraphics[height=4cm]{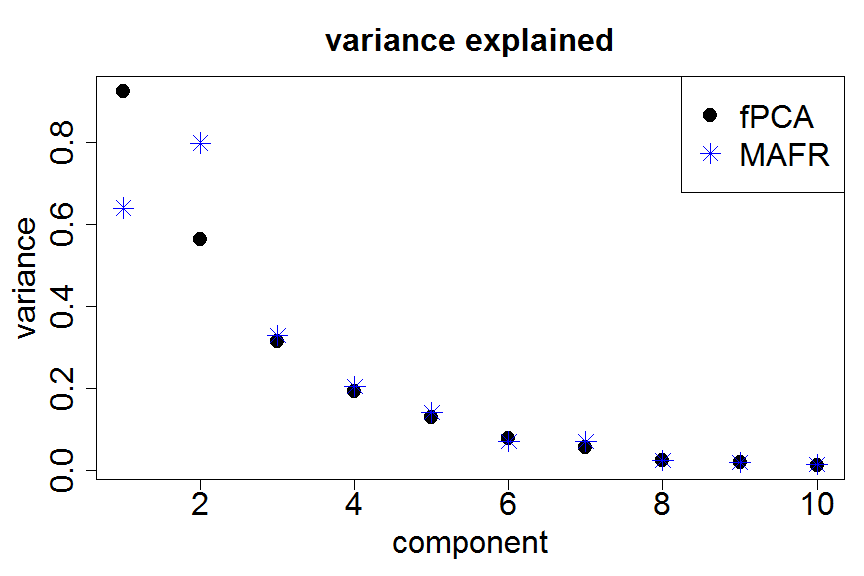} \\
\includegraphics[height=4cm]{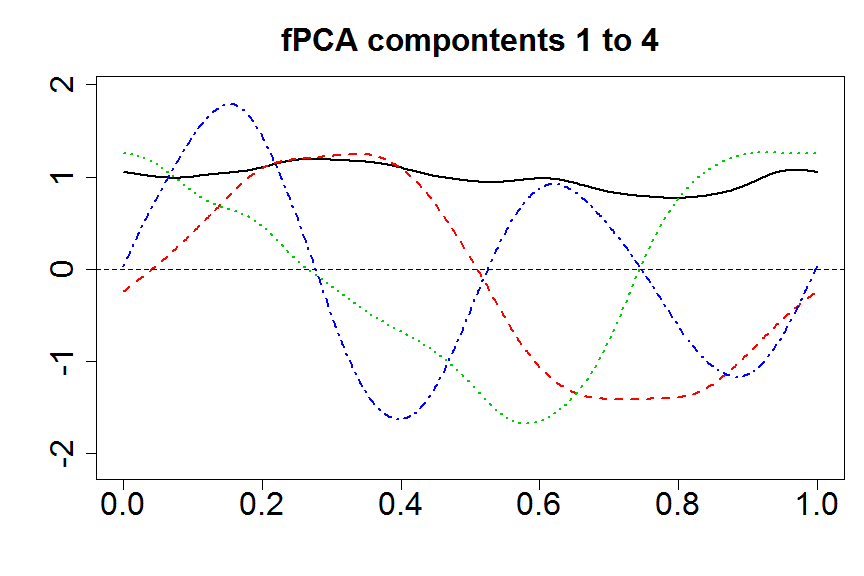} &
\includegraphics[height=4cm]{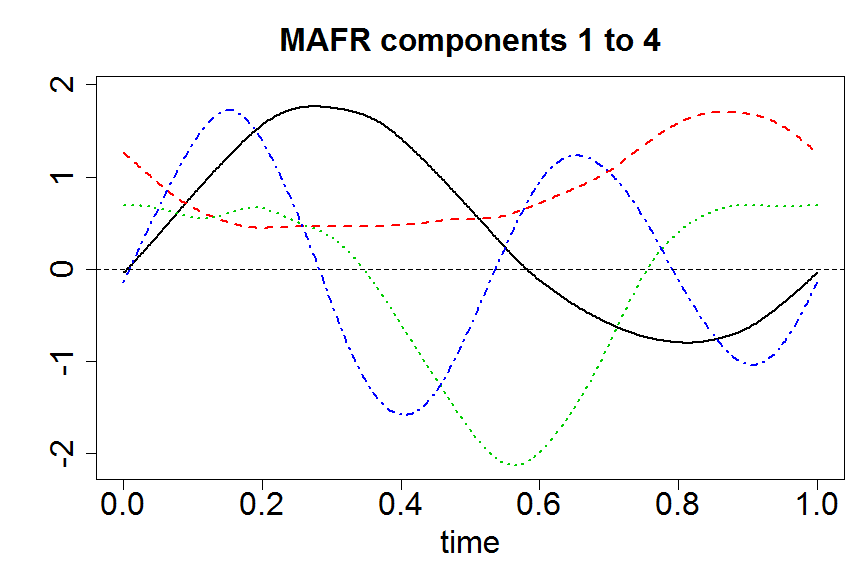} \\
\includegraphics[height=4cm]{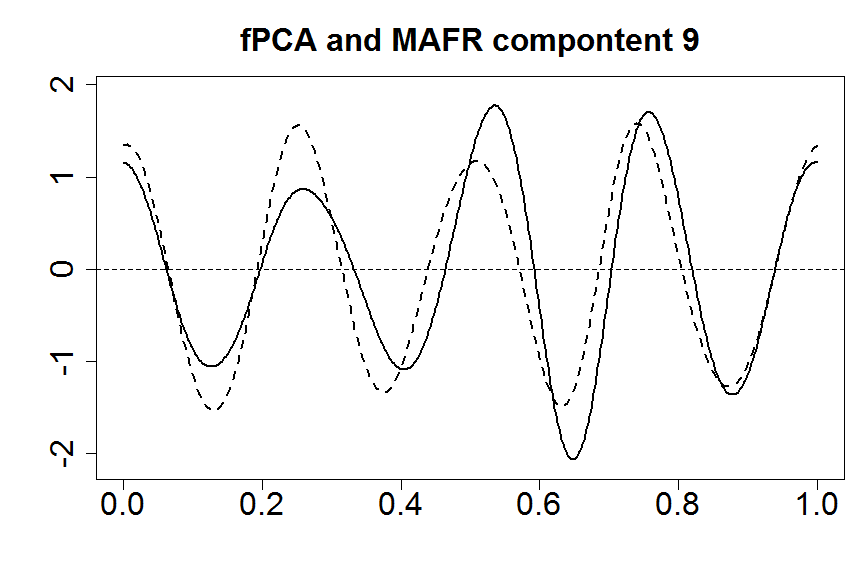} &
\includegraphics[height=4cm]{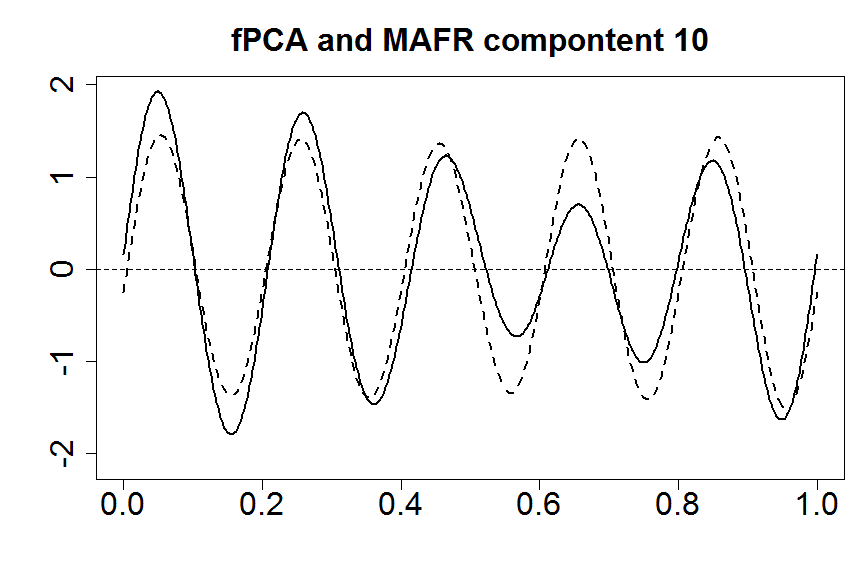}
\end{tabular}
\end{center}
\caption{Results of factor rotation based on simulated data with sinusoidal functional principal components. Top row: data (left) and variance components for fPCA (solid circles) and MAFR (stars) components.
 Second row: leading four fPCA (left) and MAFR (right) components. Bottom row: ninth (left) and tenth (right)  fPCA (solid) and MAFR (dashed) components. } \label{F:simulation}
\end{figure}

We employed a rotation based on minimizing the harmonic acceleration of the leading components and have plotted the original curves along with the first four and final two components in Figure \ref{F:simulation}. Here we see there is a distinct smoothing of the leading components. Interestingly, the final MAFR component is more purely sinusoidal -- with a higher frequency than the harmonic acceleration penalty -- than its fPCA counterpart.

These simulated data are intended as an illustrative example rather than as a quantitative investigation of the statistical properties of our method and we do not pursue a simulation here. MAFR by definition reduces the roughness of the leading principal components and can also be expected to reduce their variance. In this particular framework, it is also easy to show that rotating 25 principal components exactly recovers the original fourier basis up to changes of sign and the order of sin and cosin pairs.

\subsection{Electricity Demand Data} \label{sec:demand}
\begin{figure}
\begin{center}
\includegraphics[height=4cm]{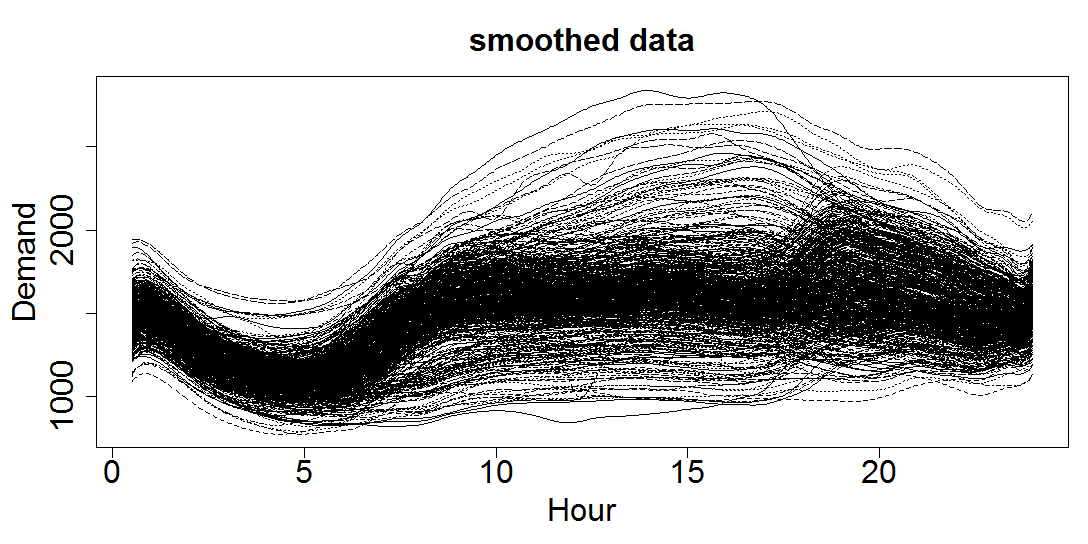}
\begin{tabular}{cc}
\includegraphics[height=4cm]{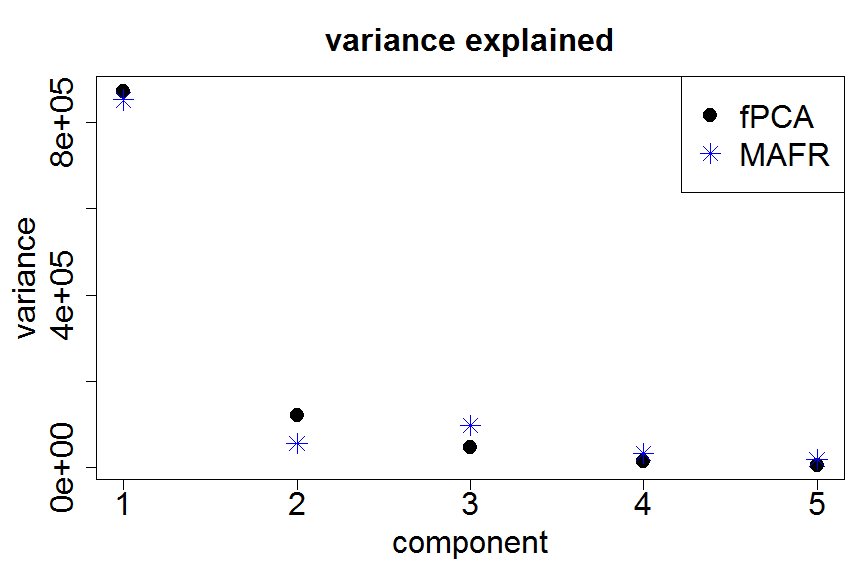} &
\includegraphics[height=4cm]{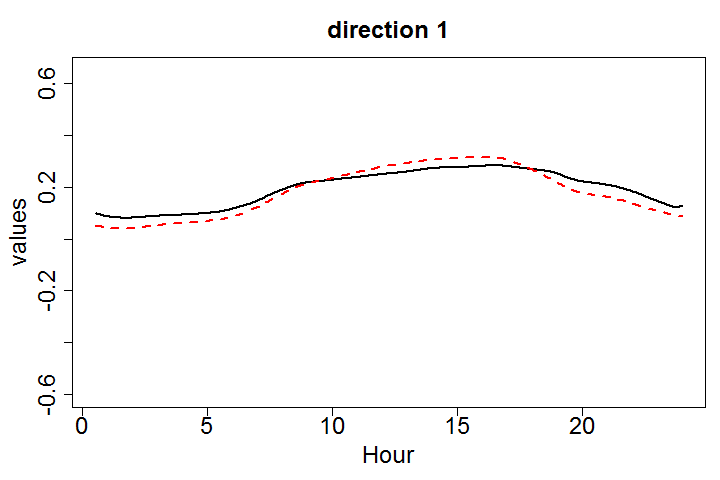} \\
\includegraphics[height=4cm]{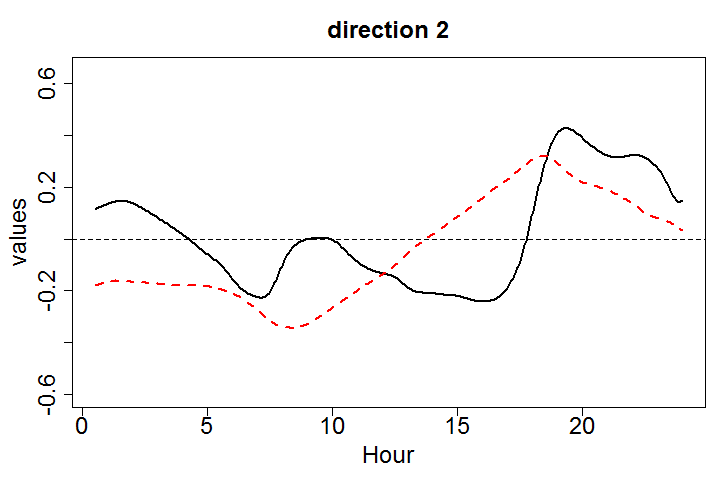} &
\includegraphics[height=4cm]{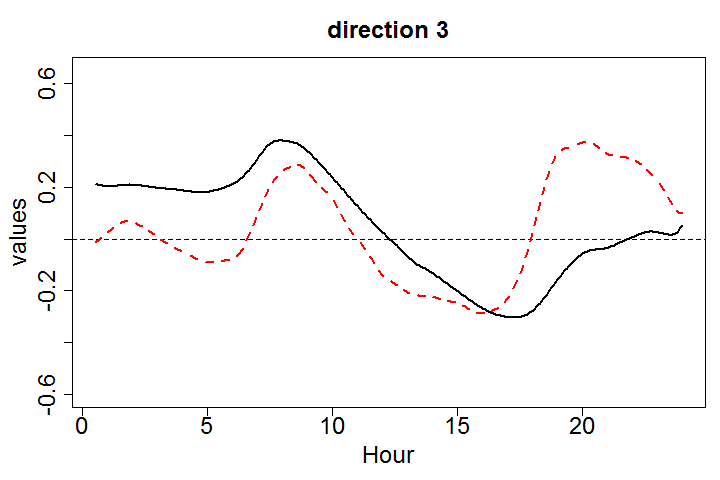} \\
\includegraphics[height=4cm]{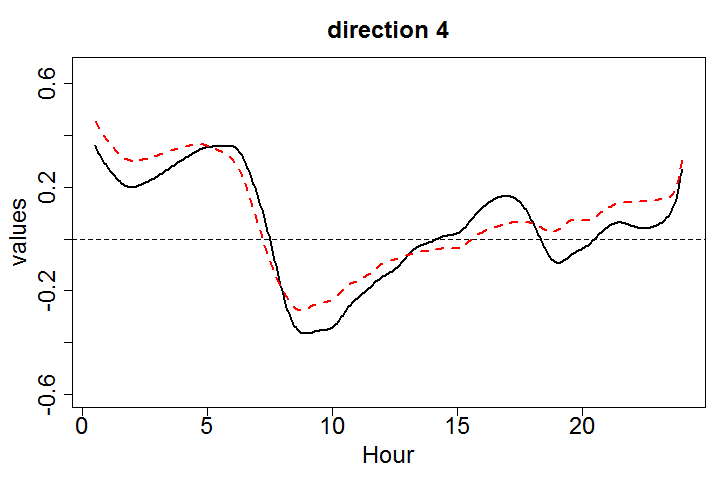} &
\includegraphics[height=4cm]{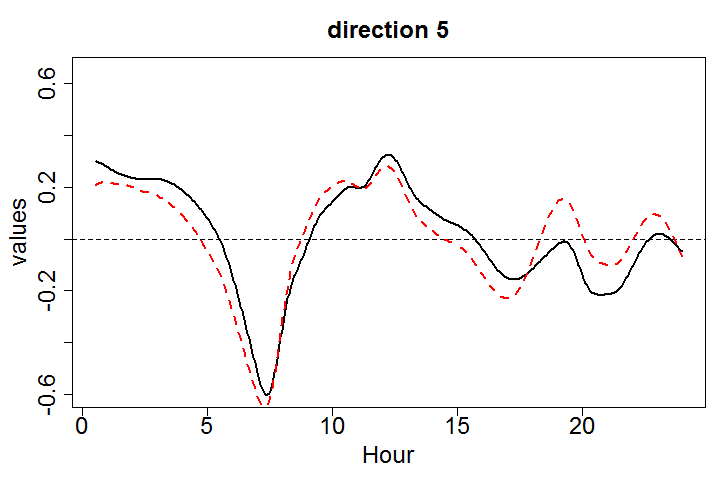}
\end{tabular}
\end{center}
\caption{Smoothed Electricity Demand Data (top) along with the fPCA (solid circles) and MAFR (stars) variance components and fPCA (black) and MAFR (red) functions. In this analysis the number of fPCA required to explain 99\% of the variation were retained.} \label{F:demand}
\end{figure}

The Electricity Demand data are obtained from the \texttt{R} package \texttt{fds} \citep{fds13}. The data comprise the half-hourly demand for electricity in Adelaide, Australia over the period 6/7/1997 to 31/3/2007. Electricity demand in Adelaide is highest in summer and winter. Interestingly the variability in electricity demand is largest when temperature is high. For further information on the Electricity Demand data the reader is referred to \citet{magnano2007generation} and \citet{magnano2008generation}. In the our analysis we restrict attention to the Monday demand for electricity and consider electricity demand as a function of time of the day. Figure \ref{F:demand} contains plots of the smoothed electricity demand versus time of the day for each of the 508 Mondays over the period of observation.

For this analysis we employed a second derivative rotation to the first five principal component directions -- these accounted for 99\% of variation in the data. The remaining plots in Figure \ref{F:demand} show the fPCA and MAFR components where a smoothing effect is evident, particularly in the second component while the later components largely retain their shape.

\section{Discussion}

This paper examines the development of factor rotations aimed specifically at providing more interpretable bases for the use of functional principal components.  Our approach here has been to find rotations that increase the smoothness of the leading principal components and we find that we are able to provide smoother leading principal components without have a large effect on the more rough components.

The proposed methods are distinct from methods which incorporate smoothing directly into a functional principal components analysis; see, for example \citet{Silverman96}. Here, we fix a subspace on which we will project our data and seek a more interpretable representation of this subspace, rather than rotating the subspace so that the original representation is smoother.

Our methods are also distinct from more classical factor rotation methods in that we target the smoothness of the factors in sequence rather than jointly.  A joint rotation criterion could be obtained by considering a weighted sum of smoothing factors:
\[
(\bbold_1,\ldots,\bbold_k) = \mbox{argmin} \sum_{j=1}^k w_j \int \left[ L \bbold_j^T \phibold(t) \right]^2 dt, \mbox{ such that } \bbold_i^T \bbold_j = I_{i=j}
\]
which is solved by the eigenvectors of $W^{1/2} P W^{1/2}$ if $W$ is a diagonal matrix with the $w_j$ on the diagonal and for which the MAFR rotation is a limiting case.  However, this then poses the problem of how to select the $w_j$ and we argue that even after deciding on a dimension, greatest attention is still paid to the leading components and these should be our focus in factor rotation.  Our methods also a variation on those of \citet{LRHF12} in defining correlation with an orthogonal basis with respect to a smoothing norm.

While we have advanced some methods designed at improving the smoothness of principal component functions, we believe that there remains considerable potential for the further development of factor rotations aimed at yielding interpretable bases specifically for functional data.

\bibliographystyle{asa}
\bibliography{references}

\end{document}